\documentclass[11pt]{article}

\usepackage{EMNLP2023}

\AtBeginDocument{%
  \providecommand\BibTeX{{%
    \normalfont B\kern-0.5em{\scshape i\kern-0.25em b}\kern-0.8em\TeX}}}

\usepackage{times}
\usepackage{latexsym}
\usepackage{microtype}
\usepackage{inconsolata}
\usepackage{float}
\PassOptionsToPackage{usenames,dvipsnames}{xcolor}
\usepackage[utf8]{inputenc} % allow utf-8 input
\usepackage[T1]{fontenc}    % use 8-bit T1 fonts
\usepackage{hyperref}       % hyperlinks
\usepackage{url}            % simple URL typesetting
\usepackage{booktabs}       % professional-quality tables
\usepackage{amsfonts}       % blackboard math symbols
\usepackage{nicefrac}       % compact symbols for 1/2, etc.
\usepackage{microtype}      % microtypography
\usepackage{tcolorbox}
\usepackage{adjustbox}
\usepackage[frozencache,cachedir=.]{minted}
\usepackage{array}
\usepackage{siunitx}
\setminted{fontsize=\scriptsize}
\definecolor{bg}{HTML}{282828}
\usepackage{pmboxdraw}
\usepackage{multirow}
\usepackage{amsmath}
\usepackage{xspace}
\usepackage{amssymb}
\usepackage{multirow}
\usepackage{multicol}
\usepackage{booktabs}
\usepackage{graphicx}
\definecolor{ForestGreen}{RGB}{34,139,34}
\definecolor{RoyalBlue}{RGB}{85,118,209}

\definecolor{Gray}{gray}{0.9}

\title{Predicting Code Coverage without Execution}

\author{Michele Tufano, Shubham Chandel, Anisha Agarwal, Neel Sundaresan, Colin Clement \\
  Microsoft \\
  Redmond, WA, USA \\
  \texttt{\{mitufano, schandel, anisagarwal, neels, coclement\}@microsoft.com}}

\begin{document}

\newcommand{\dataset}{{\sc CoverageEval}\xspace}

\newcommand{\ie}{\textit{i.e.,}~}
\newcommand{\eg}{\textit{e.g.,}~}
\newcommand{\etc}{\textit{etc.}~}
\newcommand{\etal}{\textit{et al.}~}

%Comments
\newcommand{\nb}[2]{
    \fbox{\bfseries\sffamily\scriptsize#1}
    {\sf\small$\blacktriangleright$\textit{#2}$\blacktriangleleft$}
}

\newcommand\MICHELE[1]{\textcolor{blue}{\nb{MICHELE}{#1}}}
\newcommand\COLIN[1]{\textcolor{green}{\nb{COLIN}{#1}}}
\newcommand\SHUBHAM[1]{\textcolor{red}{\nb{SHUBHAM}{#1}}}

\maketitle

\begin{abstract}
Code coverage is a widely used metric for quantifying the extent to which program elements, such as statements or branches, are executed during testing. %It provides valuable insights into the amount of code exercised by a test suite and serves as a standard measure of test suite quality. %Higher code coverage percentages indicate a lower risk of undiscovered software bugs. 
%While code coverage is not a panacea for eliminating bugs, it remains a reliable proxy metrics for assessing code quality. 
Calculating code coverage is resource-intensive, requiring code building and execution with additional overhead for the instrumentation. Furthermore, computing coverage of any snippet of code requires the whole program context. Using Machine Learning to amortize this expensive process could lower the cost of code coverage by requiring only the source code context, and the task of code coverage prediction can be a novel benchmark for judging the ability of models to understand code.
We propose a novel benchmark task called Code Coverage Prediction for Large Language Models (LLMs). We formalize this task to evaluate the capability of LLMs in understanding code execution by determining which lines of a method are executed by a given test case and inputs. We curate and release a dataset we call \dataset by executing tests and code from the HumanEval dataset and collecting code coverage information. We report the performance of four state-of-the-art LLMs used for code-related tasks, including OpenAI's GPT-4 and GPT-3.5-Turbo, Google's BARD, and Anthropic's Claude, on the Code Coverage Prediction task. Finally, we argue that code coverage as a metric and pre-training data source are valuable for overall LLM performance on software engineering tasks.

%Our ultimate goal is to leverage LLMs for predicting code coverage, offering a viable alternative to execution-based coverage in various scenarios. This approach proves advantageous when program build and execution costs are prohibitive, code coverage needs to be invoked multiple times, only code snippets are available (e.g., server-side scenarios), or errors in the project prevent complete builds. Additionally, this task serves as a novel metric for code understanding and a valuable (pre-)training objective. Training models to excel at this task could enhance their overall performance on code-related tasks.
\end{abstract}

\section{Introduction}

Software testing is an essential part of the software life-cycle which aims at detecting bugs in a program prior to shipping new versions. Code coverage is a widely used metric which estimates the quality of testing, providing some confidence that the system will operate conforming to the specified requirements. Several standards require a specific level of code coverage for software systems before they are allowed to be deployed.

% ---------- Code Example ----------------
\begin{figure}[H]
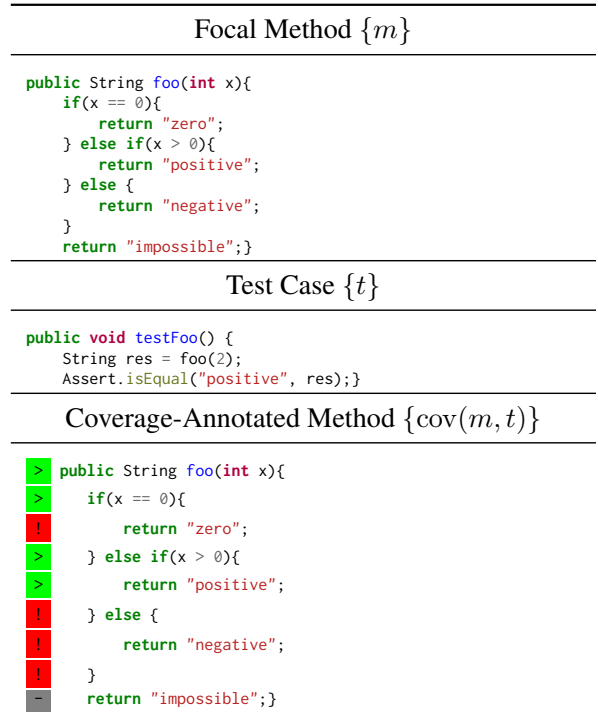

%\vspace{-0.0cm}
    \centering
\begin{adjustbox}{width=\columnwidth}
\begin{tabular}{c}
\toprule
  Focal Method $\{m\}$\\
\midrule
\begin{minipage}[t]{0.98\columnwidth}
\begin{minted}{java}
public String foo(int x){
    if(x == 0){
        return "zero";
    } else if(x > 0){
        return "positive";
    } else {
        return "negative";
    }
    return "impossible";}
\end{minted}
\end{minipage}\\

\midrule
Test Case $\{t\}$ \\
\midrule
\begin{minipage}[t]{0.98\columnwidth}
\begin{minted}{java}
public void testFoo() { 
    String res = foo(2);
    Assert.isEqual("positive", res);}
\end{minted}
\end{minipage} \\

\midrule
Coverage-Annotated Method $\{\mathrm{cov}(m,t)\}$\\
\midrule
\begin{minipage}[t]{0.98\columnwidth}
\begin{minted}[escapeinside=||]{java}
|\colorbox{green}{>}| public String foo(int x){
|\colorbox{green}{>}|    if(x == 0){
|\colorbox{red}{!}|        return "zero";
|\colorbox{green}{>}|    } else if(x > 0){
|\colorbox{green}{>}|        return "positive";
|\colorbox{red}{!}|    } else {
|\colorbox{red}{!}|        return "negative";
|\colorbox{red}{!}|    }
|\colorbox{gray}{-}|    return "impossible";}
\end{minted}
\end{minipage}
\\
\bottomrule
\end{tabular}
\end{adjustbox}
%\medskip
%\vspace{-0.0cm}
\caption{Given a focal method $m$, that is a method under test, and a test case $t$ covering that method, the code coverage obtained by $t$ on $m$ can be represented as the coverage-annotated method $\mathrm{cov}(m,t)$, where \texttt{>} represents executed statements, \texttt{!} represents statements not executed, and \texttt{-} represents unreachable code. %The lines or statements which are executed are said to be `covered', and the fraction of covered lines in a code-base is a common metric used to quantify code quality.
}
%\vspace{-0.0cm}
\label{fig:coverage}
\end{figure}
% ---------- END Code Example ----------------

For example, coverage is one of the metrics considered by the Federal Aviation Administration (FAA) for safety certification of avionic equipment, as documented in DO-178B \cite{johnson1998178b} and DO-178C \cite{rierson2017developing}. Test coverage is also a requirement in the automotive safety standard ISO 26262 Road Vehicles - Functional Safety \cite{palin2011iso}.
 
Given a focal method $m$, which is executed \textit{directly} by the test case $t$, code coverage measures the number of statements that have been executed (\ie covered) by the test $t$. Figure \ref{fig:coverage} shows an example of a focal method $m$ (method under test) tested by $t$. The coverage obtained by $t$ on $m$ is represented in the coverage-annotated method $\mathrm{cov}(m,t)$, where executed statements are marked with \colorbox{green}{>} while missed (\ie uncovered statements) with \colorbox{red}{!} and unreachable code (\ie dead code) with \colorbox{gray}{-}. From this representation, several quantitative coverage metrics can be computed, such as functional, statement, branch, and path coverage. 
%We will discuss these coverage metrics in details in the Sec \ref{sec:background}.

Code coverage is computed by instrumenting the code and running the test suite while monitoring the code execution. This process is expensive, since it requires building and executing code, especially for large software projects or when code coverage is computed multiple times. Additionally, it is not possible to measure code coverage for a snippet of code without the availability of the entire program which contains the given snippet. This situation happens when only partial code is available, for example within a commit log/diff, or when only partial code is transmitted to a server, for security and/or networking reasons.

While Large Language Models (LLMs) have gained prominence in code-related tasks and demonstrated impressive results in areas such as code generation and test generation, it remains unclear to what extent these models truly understand code execution~\cite{liu2023code}. The task of accurately determining which lines of a method are executed based on a given test case and its inputs requires a deep understanding of the underlying code execution dynamics. This motivates the need for a dedicated task, referred to as Code Coverage Prediction, which specifically evaluates the capability of LLMs in comprehending code execution. Further, a model capable of this task is independently useful as it can amortize the expensive code coverage computation process, or function in cases where normal code coverage is not possible to compute.

In this paper we formalize the Code Coverage Prediction task, with the primary objective of evaluating the capability of LLMs in understanding code execution by accurately determining which lines of a method are executed based on a given test case. To facilitate evaluation, we have curated a comprehensive dataset named \dataset, consisting of coverage-annotated methods. This dataset is created by executing tests and code from the HumanEval dataset, allowing us to collect valuable code coverage information. We have organized and made this curated dataset available on GitHub,
%and HuggingFace Datasets
 enabling researchers to explore and advance code coverage prediction techniques and LLM code understanding.

We evaluate the performance of four state-of-the-art LLMs widely employed for code-related tasks: OpenAI's GPT-4 and GPT-3.5, Google's BARD, and Anthropic's Claude. Our ultimate goal is to gain insights into the capabilities of LLMs in predicting code coverage, offering a promising alternative to execution-based coverage measurement in various scenarios. This approach proves advantageous when the costs associated with program building and execution are prohibitive, when code coverage needs to be invoked multiple times, when only code snippets are available (e.g., in server-side scenarios), or when errors in the project prevent complete builds. Additionally, this task introduces a novel metric for assessing code understanding and serves as a valuable (pre-)training objective. By training models to excel in this task, we believe we can enhance their overall performance on code-related tasks.

This paper makes the following contributions:
\begin{itemize}
\item \textit{Code Coverage Prediction Task}: We propose a novel task to assess the capability of LLMs in understanding code execution by accurately predicting executed lines of a method based on a given test case and inputs.

\item \textit{Evaluation of State-of-the-Art LLMs}: We evaluate four prominent LLMs (GPT-4, GPT-3.5, BARD, and Claude) on the Code Coverage Prediction task, providing insights into their performance and understanding of code execution.

\item \textit{Curated Dataset}: We curate a comprehensive dataset (\dataset) of coverage-annotated methods and test cases, derived from the HumanEval dataset. This dataset is openly available on GitHub\footnote{\url{https://github.com/microsoft/coverage-eval}}~\cite{CoverageDataset_github}
%\footnote{\url{https://anonymous.4open.science/r/coverage-eval-467C/}} ~\cite{CoverageDataset_github},
%and HuggingFace\footnote{\url{To-be-released}}~\cite{CoverageDataset_huggingface}, 
enabling further research and advancement in code coverage prediction techniques.
\end{itemize}

\section{Background}
\label{sec:background}

Code coverage is a measure of the degree to which a test suite exercises a software system \cite{ivankovic2019code}. Code coverage is commonly computed by means of instrumentation. This technique inserts instrumentation code in various locations within the code or binaries of the program under test, in order to monitor its execution. This inserted code provides counters to record which function or statement of the program have been executed by the test suite. Inserting these additional statements within the original code leads to execution overhead, which can be significant especially for large software programs \cite{tikir2002efficient}.

The most common coverage metric is computed at statement level, where statement refers to a syntactic unit of code (\eg assignment, invocation, assertion), often matching a single line of code. The coverage indicates whether a statement has been executed or not, and aggregated metrics can be computed at function/program level to measure the amount of statements covered by a test suite. In the example in Figure \ref{fig:coverage}, the test case $t$ executes four statements in $m$, which constitutes $\sim44$\% statement coverage for the method $m$.

Given statement coverage information, other coverage criteria and metrics can be obtained by means of static analysis. Statement coverage information regarding control structure (\eg \texttt{if-else} and \texttt{case} statements) can be used to compute branch coverage, which measure how many logical branches in the program have been executed. In the example in Figure \ref{fig:coverage} only one branch is executed (\ie \texttt{else if (x > 0)} ), while the other two branches are missed by the test case $t$.

In the remainder of this paper we will focus on statement coverage, from which other coverage criteria can be obtained.

\section{Code Coverage Prediction Task}
\label{sec:problem}
Given a method under test (focal method) $m$, composed of $n$ statements $S_m = s_1, s_2, \dots, s_n$, and a test case $t$ which exercises the method $m$, the coverage-annotated focal method $\mathrm{cov}(m,t)$ is composed of a sequence of $n$ statements $S_{m}^{t} = s_1^*, s_2^*, \dots, s_n^*$, where each statement  $s_i^*$ represents the coverage-annotated statement of $s_i$ in $m$. Specifically, $s_i^*$ is marked with one of the three possible coverage symbols $c \in \{>, !, - \}$, where the symbol $>$ identifies statements that have been executed by $t$, the symbol $!$ identifies statements that have been missed by $t$, and the symbol $-$ identifies statements that are unreachable. This defines a sequence of $n$ coverage symbols $C_m^t = c_1, c_2, \dots, c_n$, where $c_i \in \{>, !, - \}$. %This sequence can be combined with the original sequence of statements $S_m = s_1, s_2, \dots, s_n$, to obtain the coverage-annotated sequence of statements $S_m^t = s_1^*, s_2^*, \dots, s_n^*$ comprising the coverage $\mathrm{cov}(m,t)$.

We define the Code Coverage Prediction Task as the problem of predicting the coverage-annotated sequence of statements $S_m^t$ given the focal method $m$ and a test case $t$. Formally, this problem can be defined in terms of inputs and expected output:

\textbf{Input}
        \begin{itemize}
        \item Focal Method: $m$
        \item Test Case: $t$
    \end{itemize}
    
\textbf{Output}
    \begin{itemize}
        \item $S_m^t = s_1^*, s_2^*, \dots, s_n^*$ \\
        or
        \item $C_m^t = c_1, c_2, \dots, c_n$
    \end{itemize}

Specifically, the output can be either the coverage-annotated sequence of statements $S_m^t$, or the sequence of coverage symbols $C_m^t$, which can then combined with the original sequence of statements $S_m = s_1, s_2, \dots, s_n$, to obtain the coverage-annotated sequence of statements $S_m^t = s_1^*, s_2^*, \dots, s_n^*$ comprising the coverage $\mathrm{cov}(m,t)$. This final step is performed by aligning the two sequences and obtaining $s_i^* = c_i + s_i$, where the $+$ operation refers to string concatenation. 

Let us take as example the focal method $m$ and test case $t$ in Figure \ref{fig:coverage}. The model is expected to predict either the coverage-annotated sequence of statements $S_m^t$ or the sequence of coverage symbols: \texttt{> > ! > > ! ! ! -}. 
%This sequence of symbols is then aligned with the original sequence of statements, and the coverage-annotated focal method $c$ is obtained by concatenating each coverage symbol $c_i$ to the corresponding statement $s_i$.\COLIN{Is this still the case or do we just directly predict the annotated statements?} \MICHELE{Good point. Clarified this.}

\subsection{Coverage Prediction for Pre-Training}

We propose that the code coverage prediction task introduced in our paper can serve as a valuable pre-training task for LLMs focused on code generation. While current pre-training tasks, such as Masked Language Modeling (MLM) help models understand code syntax and semantics by analyzing vast amounts of raw text representing code, our proposed task enables the model to learn about code execution, which is not technically discoverable by source code text alone.

To accomplish this pre-training, we suggest augmenting the training data with extensive coverage logs obtained from Continuous Integration/Continuous Deployment (CI/CD) pipelines. These logs contain valuable information about code coverage from regression tests executed during pull requests or commits.

By exposing the models to these coverage logs during pre-training, they can learn to associate test cases and inputs with the specific lines of code that are executed. This pre-training approach enhances the models' understanding of how different parts of the code are exercised by various test scenarios. Consequently, the models can acquire a deeper comprehension of the relationships between inputs, tests, and code execution, leading to improved code generation capabilities.

Integrating coverage prediction as a pre-training task could enable models to learn from real-world test scenarios, capturing the nuances of code execution in practical settings. This real-world exposure should enhances the models' ability to generate code that aligns with actual testing practices.

Furthermore, incorporating coverage prediction as a pre-training task opens up possibilities for transfer learning. Models pre-trained on coverage prediction can be fine-tuned on downstream tasks, such as bug detection or test case generation, where understanding code execution is crucial. The models' pre-existing knowledge of code coverage can provide a solid foundation for these related tasks, potentially improving their overall performance.

\section{\dataset Dataset}
In addition to proposing the code coverage prediction task, this paper also introduces \dataset, a dataset specifically designed for evaluating LLMs on this task. This section outlines the process of curating this dataset, which begins with the HumanEval dataset \cite{chen2021codex}. By executing test cases from the HumanEval dataset, we gather code coverage information. To create \dataset, we parse the code coverage logs generated during the execution of the test cases. This parsing step enables us to extract the relevant coverage annotations. We then carefully structure and export the dataset in a format that facilitates its use and evaluation by researchers and practitioners alike.

By curating this dataset, we aim to provide a standardized benchmark for evaluating LLMs on the code coverage prediction task. The availability of \dataset enables researchers to explore and advance code understanding, fostering innovation and enabling the development of more effective models.

\subsection{HumanEval}
The HumanEval dataset consists of 164 hand-written problems and their code solutions, where each problem is a programming task involving language comprehension, reasoning, algorithms and/or simple mathematics \cite{chen2021codex}. Each code solution in the dataset includes a function signature, a docstring containing the problem description, a function body, and several unit tests. We extend the HumanEval dataset to include coverage, calculated using the function body and the respective unit tests.

\subsection{Coverage Analysis}
In this section, we describe the steps taken to analyze the code coverage on the HumanEval dataset and create our \dataset dataset.

Each code solution in the HumanEval dataset is accompanied by a single test case, which includes multiple asserts designed to test the correctness of the code solution based on the given problem's functional requirements. These asserts cover various inputs, scenarios, and code statements/branches. To enhance the dataset and increase the complexity of each data point, we split the single test case into multiple test cases, each containing a single assert. This splitting process allows us to generate additional method-test pairs, as well as making each data point more challenging. The original test case may cover most of the lines and branches in the method, but each individual assert covers only a subset of them.

By performing this split, we create a more diverse set of method-test pairs within the dataset. Each individual test case invokes the focal method once and covers a subset of the statements and branches within the method. This enables us to evaluate the LLMs' ability to predict code coverage at a more granular level, going beyond the overall coverage of the method. It also adds complexity to the task, as predicting coverage for each assert requires a deeper understanding of the code and its potential execution paths.

Subsequently, we execute the extracted test cases individually with \texttt{pytest}. During the execution, we also enable the coverage computation using \texttt{coverage.py}. To do so, we run the following command: \texttt{coverage run -m pytest <test\_name>} where \texttt{<test\_name>} is each individual test in the dataset.

Next, for each test case $t$, we analyze the corresponding coverage report obtained by the test execution in order to extract the annotated coverage $\mathrm{cov}(m,t)$. The coverage report marks each source code line in the file with coverage information, specifying whether the statement has been executed or not.

We automatically parse this report and extract the corresponding annotated coverage $\mathrm{cov}(m,t)$. At the end of this process, we obtained a dataset where each data point is formed by a triplet $d = \{ m, t, \mathrm{cov}(m,t) \}$.

\subsection{Data Format}
The \dataset dataset maintains the structure of the HumanEval dataset, with the addition of coverage information for each test. Each record corresponds to a unique problem and contains the following fields:

\begin{itemize}
  \item Problem ID: A unique ID for the problem
  \item Problem: The name of the method written to solve the problem
  \item Method: The method contents, including a function signature, a docstring with the details of the problem, and the function body.
  \item Tests: A list of unit tests for the problem. Each item in the list includes the unique ID of the test and the code of the test. We have also added coverage information for each test in the following two forms:
    \begin{enumerate}
      \item Coverage: The code of the method, with each line annotated with \colorbox{green}{>}, \colorbox{red}{!} or \colorbox{gray}{-} for code that is executed, missed or unreachable by the given test.
    \item Coverage Sequence: A list of equal length to the number of lines in the method, where each value in the list is \colorbox{green}{>}, \colorbox{red}{!} or \colorbox{gray}{-}, depending on the status of the respective line of code in the method.
    \end{enumerate}
\end{itemize}

Figure \ref{fig:coval_data_example} (Appendix) shows a sample record from the \dataset dataset. \dataset is available to the public via GitHub~\cite{CoverageDataset_github}.
%and HuggingFace~\cite{CoverageDataset_huggingface}.

Table \ref{tab:dataset} reports the statistics for the \dataset dataset in terms of number of problems, code solutions, tests, and coverage symbols. The discrepancy between number of problems and solutions is explained by the fact that some problems have multiple solutions. It is also worth noting that while our dataset currently does not contain any unreachable code (-), we have proactively considered the potential presence of unreachable code while designing the task.

\begin{table}[]
\resizebox{0.5\textwidth}{!}{
\begin{tabular}{@{}cccccc@{}}
\toprule
\multirow{2}{*}{Problems} & \multirow{2}{*}{Solutions} & \multirow{2}{*}{Tests} & \multicolumn{3}{c}{Coverage Symbols}                     \\ \cmidrule(l){4-6} 
                          &                            &                        & Executed (\textgreater{}) & Missed (!) & Unreachable (-) \\ \midrule
                158       &                 164     &           1160            &                      20037 & 1734      & 0               \\
%\multicolumn{1}{l}{}      & \multicolumn{1}{l}{}       &                        &                           &            &                 \\ 
\bottomrule
\label{tab:dataset}
\end{tabular}
}
\vspace{-0.5cm}
\caption{\dataset statistics.}
\label{tab:dataset}
\vspace{-0.5cm}
\end{table}

\section{Evaluating LLMs}

In this section, we present our evaluation of state-of-the-art Language Models (LLMs) for the proposed task of Code Coverage Prediction. We selected four highly regarded LLMs that are not only popular for code generation but also widely used for other Natural Language (NL) tasks. The LLMs we employed for this evaluation are OpenAI's GPT-4 and GPT-3.5, Google's BARD, and Anthropic's Claude.

GPT-3.5 \cite{brown2020language} and GPT-4 \cite{openai2023gpt4} are large language models developed by OpenAI which are Transformer-style models \cite{DBLP:journals/corr/VaswaniSPUJGKP17} pre-trained to predict the next token in a document. Both models were then fine-tuned using Reinforcement Learning from Human Feedback (RLHF) \cite{christiano2017deep}. GPT-4 improves over the predecessor by accepting as input both images and text (multimodal model) and producing text as output. BARD is a conversational AI developed by Google based on LaMDA\cite{thoppilan2022lamda} a Transformer-based language models trained on dialogue \cite{adiwardana2020towards}. Anthropic Claude is a 52-billion-parameter LLM developed by Anthropic. Claude was pretrained on a large text corpus and finetuned with "RL from AI Feedback" (RLAIF), where AI feedback are steered by a small set of principles drawn from a "constitution" defined by humans \cite{bai2022constitutional}.

\subsection{Experimental Design}
%\SHUBHAM{Updated the datapoints. Looks good to me.} \MICHELE{Great!}
When evaluating the LLMs on the code coverage prediction task, we designed the experiments to assess their performance on non-trivial coverage sequences while progressively providing more information and examples.

First, we filtered out data points $d = \{ m, t, \mathrm{cov}(m,t) \}$ where the coverage sequence is \textit{trivial} consisting exclusively of the symbol \colorbox{green}{>}. These cases represent methods with no branches or where the test case covers every statement in the focal method. Although these data points are included in the \dataset dataset, we excluded them from this specific evaluation. The subset of data points containing only trivial symbols is reported in our online appendix. It's important to note that no data points in the dataset has a coverage sequence consisting solely of \colorbox{red}{!} or \colorbox{gray}{-} symbols. After this filtering step, we were left with 478 data points on which we evaluated the LLMs.

The prompt used to evaluate the LLMs was designed to include the following sections:
\begin{itemize}
\item System NL prompt: a prompt providing a natural language description of the task, aimed at conveying the task to the LLM.
\item Examples: zero, one, or multiple examples of the task.
\item Focal Method $m$ and Test Case $t$.
\end{itemize}

In terms of the System NL prompt, our evaluation involved experimenting with various prompts and descriptions. We achieved the most favorable outcomes by utilizing a system prompt that emulates a terminal environment (e.g., python terminal). Within this prompt, we instructed the LLM to generate the code coverage output based on a given test case and method. For OpenAI models, we included this prompt in the specific system prompt section, while for BARD and Claude, we incorporated it as the initial part of the prompt.

To comprehensively assess the LLMs' performance, we conducted evaluations using different numbers of examples for the code coverage prediction task. Specifically, we employed zero-shot, one-shot, and multi-shot prompting approaches. This allowed us to examine the impact of example availability on the models' performance and their ability to generalize the task across various methods.

When selecting examples for evaluating coverage on a particular method $m_i$, we took care to prevent data leakage and encourage the LLMs to generalize their predictions to other methods. To achieve this, we randomly sampled a data point $\{m_j, t, \mathrm{cov}(m,t)\}$ where $m_j \neq m_i$ when providing examples.

\begin{figure}[H]
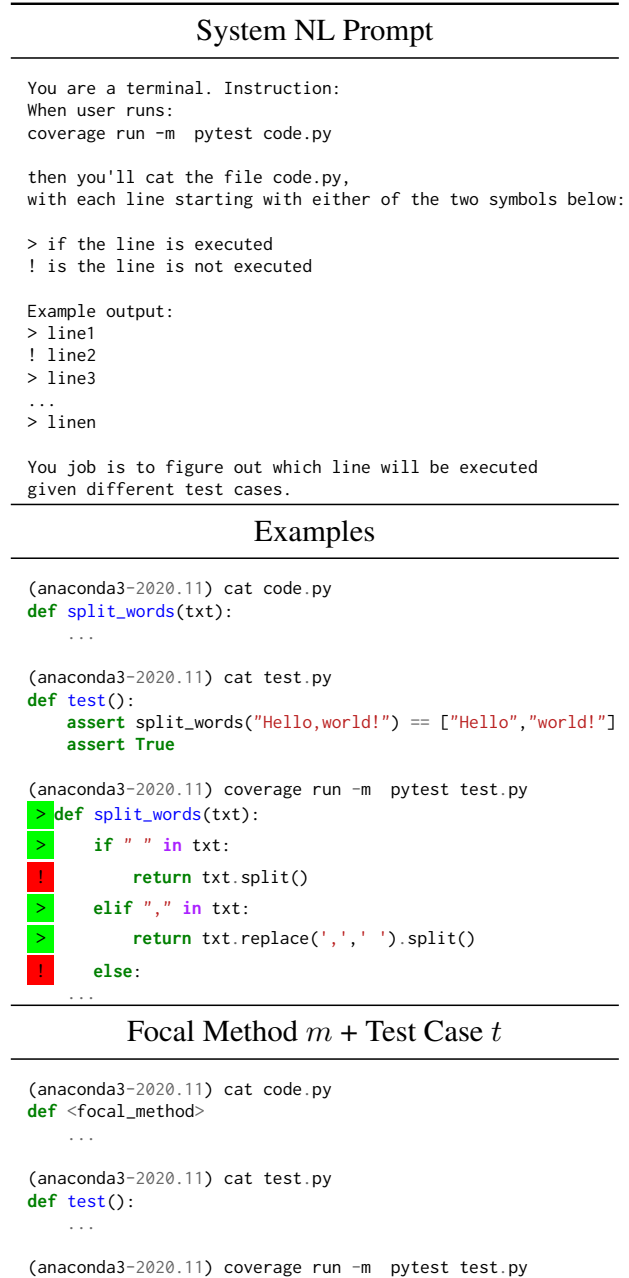

\vspace{-0.0cm}
    \centering
\begin{adjustbox}{width=0.5\textwidth}
\begin{tabular}{c}
\toprule
  System NL Prompt\\
\midrule
\begin{minipage}[t]{0.45\textwidth}
\begin{minted}{text}
You are a terminal. Instruction:
When user runs:
coverage run -m  pytest code.py

then you'll cat the file code.py, 
with each line starting with either of the two symbols below:

> if the line is executed
! is the line is not executed

Example output:
> line1
! line2
> line3
...
> linen

You job is to figure out which line will be executed 
given different test cases.
\end{minted}
\end{minipage}\\
\midrule
Examples \\
\midrule
\begin{minipage}[t]{0.45\textwidth}
\begin{minted}[escapeinside=||]{python}
(anaconda3-2020.11) cat code.py
def split_words(txt):
    ...

(anaconda3-2020.11) cat test.py
def test():
    assert split_words("Hello,world!") == ["Hello","world!"]
    assert True

(anaconda3-2020.11) coverage run -m  pytest test.py
|\colorbox{green}{>}|def split_words(txt):
|\colorbox{green}{>}|    if " " in txt:
|\colorbox{red}{!}|        return txt.split()
|\colorbox{green}{>}|    elif "," in txt:
|\colorbox{green}{>}|        return txt.replace(',',' ').split()
|\colorbox{red}{!}|    else:
    ...
\end{minted}
\end{minipage} \\
\midrule
Focal Method $m$ + Test Case $t$\\
\midrule
\begin{minipage}[t]{0.45\textwidth}
\begin{minted}[escapeinside=||]{python}
(anaconda3-2020.11) cat code.py
def <focal_method>
    ...

(anaconda3-2020.11) cat test.py
def test():
    ...

(anaconda3-2020.11) coverage run -m  pytest test.py

\end{minted}
\end{minipage}
\\
\bottomrule
\end{tabular}
\end{adjustbox}
%\medskip
\vspace{-0.0cm}
\caption{Code Coverage Prediction Task Prompt: (i) System NL Prompt instruct the LLM to operate as in a terminal environment; (ii) zero, one, or multiple examples of the coverage prediction task may be shown; (iii) the current focal method $m$ and test case $t$ are provided}
\vspace{-0.0cm}
\label{fig:prompt}
\end{figure}

Finally, the prompt provides a focal method $m$ and a corresponding test case $t$ for which we expected the model to predict the code coverage. Figure \ref{fig:prompt} shows an example of the prompt we designed.

Inference is performed on all the LLMs with temperature and topp set to 0, and generating one sample.

% ---------- Code Example ----------------
% ---------- END Code Example ----------------

\subsection{Evaluation Metrics}
In this section we describe the evaluation metrics.

Given the method $m$, the test case $t$, and the sequence of coverage symbols $C_m^t = c_1, c_2, \dots, c_n$, where $c_i \in \{>, !, - \}$, the model generates a predicted sequence of coverage symbols $\hat{C}_m^t = \hat{c}_1, \hat{c}_2, \dots, \hat{c}_n$. We consider the following metrics to evaluate the performances of our proposed approach.

\begin{table*}[]
\resizebox{\textwidth}{!}{%
\begin{tabular}{@{}lccccccccc@{}}
\toprule
\multirow{2}{*}{Model} & \multicolumn{3}{c}{zero-shot}                                 & \multicolumn{3}{c}{one-shot}                                 & \multicolumn{3}{c}{multi-shot}                                 \\ \cmidrule(l){2-10} 
                       & \multicolumn{1}{c}{Match} & \multicolumn{1}{c}{Stmt} & Branch & \multicolumn{1}{c}{Match} & \multicolumn{1}{c}{Stmt} & Branch & \multicolumn{1}{c}{Match} & \multicolumn{1}{c}{Stmt} & Branch \\ \midrule
OpenAI GPT-4   (gpt-4)        & \textbf{25.75} & \textbf{84.47} & \textbf{20.16} &\textbf{ 22.85} & \textbf{90.71} & \textbf{22.65} & \textbf{30.04} & \textbf{90.5} & \textbf{22.5} \\
OpenAI GPT-3.5 (gpt-3.5-turbo)         &0 & 39.87 & 8.33 & 8.17 & 76.53 & 17.17 & 11.03 & 82.29 & 17.9 \\
Google BARD (text-bison-001)            &0 & 81.27 & 17.21 & 1.87 & 86.93 & 19.63 & 21.56 & 85.66 & 20.52 \\
Anthropic Claude (claude-1.3)       &3.9 & 84.47 & 20.07 & 4.83 & 83.21 & 19.16 & 6.88 & 55.7 & 12.23 \\
 \bottomrule
\end{tabular}
}
\caption{LLMs performances on the Code Coverage Prediction Task. The table reports the percentages of predicted coverage sequences that match the ground truth (Match), the percentage of correct coverage symbols for statements (Stmt), and specifically for branches (Branch). Evaluation performed for zero-shot, one-shot, and multi-shot.}
\label{tab:results}
\end{table*}

\subsubsection{Perfect Sequence Match}
The perfect sequence match metric counts the number of times that the predicted sequence $\hat{C}_m^t$ exactly matches (symbol-by-symbol) the target coverage sequence $C_m^t$. This represents the case where the model predicts the coverage with perfect accuracy for all the statements and branches.

\subsubsection{Statement Correctness}
The statement correctness metric measures the percentage of statements for which the execution prediction is correct. This is equivalent to the percentage of symbols in the predicted sequence that match the target sequence.

\subsubsection{Branch Correctness}
The branch correctness metric measures the percentage of branch-specific statements for which the execution prediction is correct.
The branch correctness only considers the symbols associated with branch statements. It measures the percentage of symbols in the predicted sequence (associated with branches) that match the symbols in the target sequence.

\section{Results}

Table \ref{tab:results} presents the performance of different LLMs on the Code Coverage Prediction task. The table showcases the percentage of predicted coverage sequences that match the ground trught (Match), the percentage of correct coverage symbols for all the statements (Stmt), and the percentage of correct coverage symbols when only considering branch statements (Branch). Evaluation performances are computed using zero-shot, one-shot, and multi-shot prompting.

OpenAI GPT-4 demonstrates the highest performance on this task, achieving 24.75\% exact match with zero-shot prompting and improving to 30\% with multi-shot prompting, where up to 6 examples are provided in the prompt. Notably, the other LLMs achieve low exact matches with zero-shot prompting (between 0 and 4\%), suggesting that these foundational models may not have been exposed to coverage logs during their training or that. The second best-performing model is Google BARD, with an exact sequence match reaching 21.5\% with multi-shot prompting.

Regarding the percentage of correct coverage statements (see Stmt), most models demonstrate improvement as more examples are included in the prompt. OpenAI GPT-4 obtain the overall best scores between 84\% and 90\% of statement correctness.

When considering only statements involved in branches (\eg \texttt{if-else}, \texttt{while}), it becomes evident that there is a significant drop in correct predictions. In fact, the best performing model, OpenAI GPT-4, accurately predicts a modest 22\% of these symbols when one- and multi-shot is used for prompting. It is important to note that this subset of statements, which are intricately connected to branches, presents a greater challenge for evaluation because the LLM must reason about the boolean conditions that determine which branch is covered. Consequently, accurately predicting coverage symbols within this context requires the model to possess a profound understanding of the conditional logic that guides program execution.

Despite the surprisingly strong results of OpenAI GPT-4 on the Code Coverage Prediction task, it should be noted that the model still fails to generate the correct coverage for more than 70\% of the method-test pairs in the \dataset dataset. This emphasizes that LLMs have a long way to go in developing a deep understanding of code execution.

We believe that in order to enhance code generation results, these LLMs should gain a comprehensive understanding of code execution under different inputs and test cases. Therefore, we assert that our dataset and proposed task can contribute to the advancement of LLMs towards this goal.

\section{Discussion\& Applications}

LLMs trained to excel on the Code Coverage Prediction task could offer a promising alternative to traditional execution-based code coverage measurement in various scenarios. In this section, we discuss several use case scenarios where this approach can be valuable and beneficial.

\subsection{Expensive Build \& Execution}

For large software projects with millions of lines of code and numerous dependencies, the build and execution process can be time-consuming and expensive. In such cases, developers may want to analyze the code coverage obtained by newly written tests without waiting for the lengthy build phase. By leveraging LLMs trained on the Code Coverage Prediction task, developers can predict the coverage obtained by the new tests on existing methods without the need to build the entire project or execute the tests. This enables developers to quickly assess whether additional tests are required to cover missed lines or branches in the methods, saving valuable time and resources.

\subsection{Limited Code Availability}

Traditional code coverage computation requires the complete source code of the codebase to be available for instrumentation and execution. However, there are scenarios where only a partial view of the code is accessible, making code coverage computation impossible using traditional methods. 

In cases where limited code availability poses a challenge, the Code Coverage Prediction approach can be employed. For example, when utilizing an AI code generation service from an IDE, developers may transmit only a partial view of the code to the server where the AI model resides. In this scenario, the server can use the proposed approach to predict the code coverage of the AI-generated test cases on the given method. This enables estimation of the code coverage without the need for the entire codebase, addressing privacy concerns and network limitations. The predicted code coverage can then be used to make informed decisions, such as generating additional tests if coverage is insufficient or transmitting the generated tests to the user if coverage is satisfactory.

\subsection{Live Coverage}

Live Unit Testing, integrated into various IDEs, allows developers to receive real-time feedback on the impact of code changes on existing tests and identifies whether newly added or modified code is covered by existing tests. In this scenario, the Code Coverage Prediction approach can be applied by replacing the actual execution of test cases with an AI inference call to predict the coverage on the modified or newly added methods. This provides developers with immediate feedback on code coverage without the need for executing the entire test suite. By utilizing LLM-based models for code coverage prediction, developers can streamline the testing process and receive timely insights into the coverage of their code changes.

\section{Conclusion}
In this paper, we introduced the novel task of Code Coverage Prediction, which aims to assess the capabilities of Large Language Models (LLMs) in understanding code execution by accurately predicting the lines of code that are executed based on given test cases. We curated a comprehensive dataset named \dataset, consisting of coverage-annotated methods derived from the HumanEval dataset. This dataset enables researchers to explore and advance code coverage prediction techniques and LLM code understanding.

We evaluated the performance of four state-of-the-art LLMs, namely OpenAI's GPT-4 and GPT-3.5, Google's BARD, and Anthropic's Claude, on the Code Coverage Prediction task. The results demonstrated that GPT-4 achieved the highest performance, with 10.46\% exact match with zero-shot prompting and 24.48\% with multi-shot prompting. However, none of the models, including GPT-4, achieved high accuracy in predicting code coverage, indicating that LLMs still have a long way to go in developing a deep understanding of code execution.

The Code Coverage Prediction task serves as a valuable metric for assessing code understanding and can potentially contribute to the enhancement of LLMs' overall performance on code-related tasks. By training models to excel in this task, we can improve their ability to comprehend code execution dynamics, which is crucial for tasks such as code generation and test generation.

%\bibliographystyle{ACM-Reference-Format}
%\bibliography{main}

%\bibliographystyle{IEEEtran}
% \bibliographystyle{ACM-Reference-Format}
\bibliography{main}
\bibliographystyle{acl_natbib}

\newpage

\appendix

\section{\dataset Example}

% ---------- Data Example ----------------
\begin{figure}[H]
\vspace{-0.0cm}
    \centering
\begin{adjustbox}{width=0.5\textwidth}
\begin{tabular}{c}
\toprule
  Problem: rounded\_avg\\
\midrule
\small{ID = 104}\\
\begin{minipage}[t]{0.45\textwidth}
\begin{minted}{python}
def rounded_avg(n, m):
    """SYNTH You are given two positive integers n and m,
    and your task is to compute the average of the integers
    from n through m (including n and m). 
    Round the answer to the nearest integer
    and convert that to binary.
    If n is greater than m, return -1.
    Example:
    rounded_avg(1, 5) => "0b11"
    rounded_avg(7, 5) => -1
    rounded_avg(10, 20) => "0b1111"
    rounded_avg(20, 33) => "0b11010"
    """
    if m < n:
        return -1
    summation = 0
    for i in range(n, m+1):
        summation += i
    return bin(round(summation/(m - n + 1)))
\end{minted}
\end{minipage}\\
\midrule
Test Cases\\
\midrule
\begin{minipage}[t]{0.45\textwidth}
\begin{minted}{python}
def test_658():
    assert rounded_avg(185,546) == "0b101101110"
    assert True

def test_659():
    assert rounded_avg(362,496) == "0b110101101"
    assert True

def test_660():
    assert rounded_avg(560,851) == "0b1011000010"
    assert True

\end{minted}
\end{minipage} \\
\midrule
Coverage-Annotated Method $\{test\_id = 660\}$\\
\midrule
\begin{minipage}[t]{0.45\textwidth}
\begin{minted}[escapeinside=||]{python}
|\colorbox{green}{>}| def rounded_avg(n, m):
|\colorbox{green}{>}|    if m < n:
|\colorbox{red}{!}|         return -1
|\colorbox{green}{>}|     summation = 0
|\colorbox{green}{>}|     for i in range(n, m+1):
|\colorbox{green}{>}|         summation += i
|\colorbox{green}{>}|     return bin(round(summation/(m - n + 1)))

\end{minted}
\end{minipage}\\
\midrule
Coverage Sequence $\{test\_id = 660\}$\\
\midrule
\begin{minipage}[t]{0.45\textwidth}
\begin{minted}[escapeinside=||]{python}
[|\colorbox{green}{>}|, |\colorbox{green}{>}|, |\colorbox{red}{!}|, |\colorbox{green}{>}|, |\colorbox{green}{>}|, |\colorbox{green}{>}|, |\colorbox{green}{>}|]
\end{minted}
\end{minipage}
\\
\bottomrule
\end{tabular}
\end{adjustbox}
%\medskip
\vspace{-0.0cm}
\caption{Example record from the \dataset dataset. This record is for the rounded\_avg problem. We have shown 3 of the unit tests, as well as sample coverage annotation data from one unit test. where \texttt{>} represents executed statements, \texttt{!} missing statements, and \texttt{-} unreachable code.}
\vspace{-0.0cm}
\label{fig:coval_data_example}
\end{figure}

\newpage

\section{Deployed Systems}
We deploy our approach in two systems covering some of the use cases described in the paper.

\subsection{System A - Live Coverage}
\begin{figure*}
\includegraphics[width=0.95\textwidth]{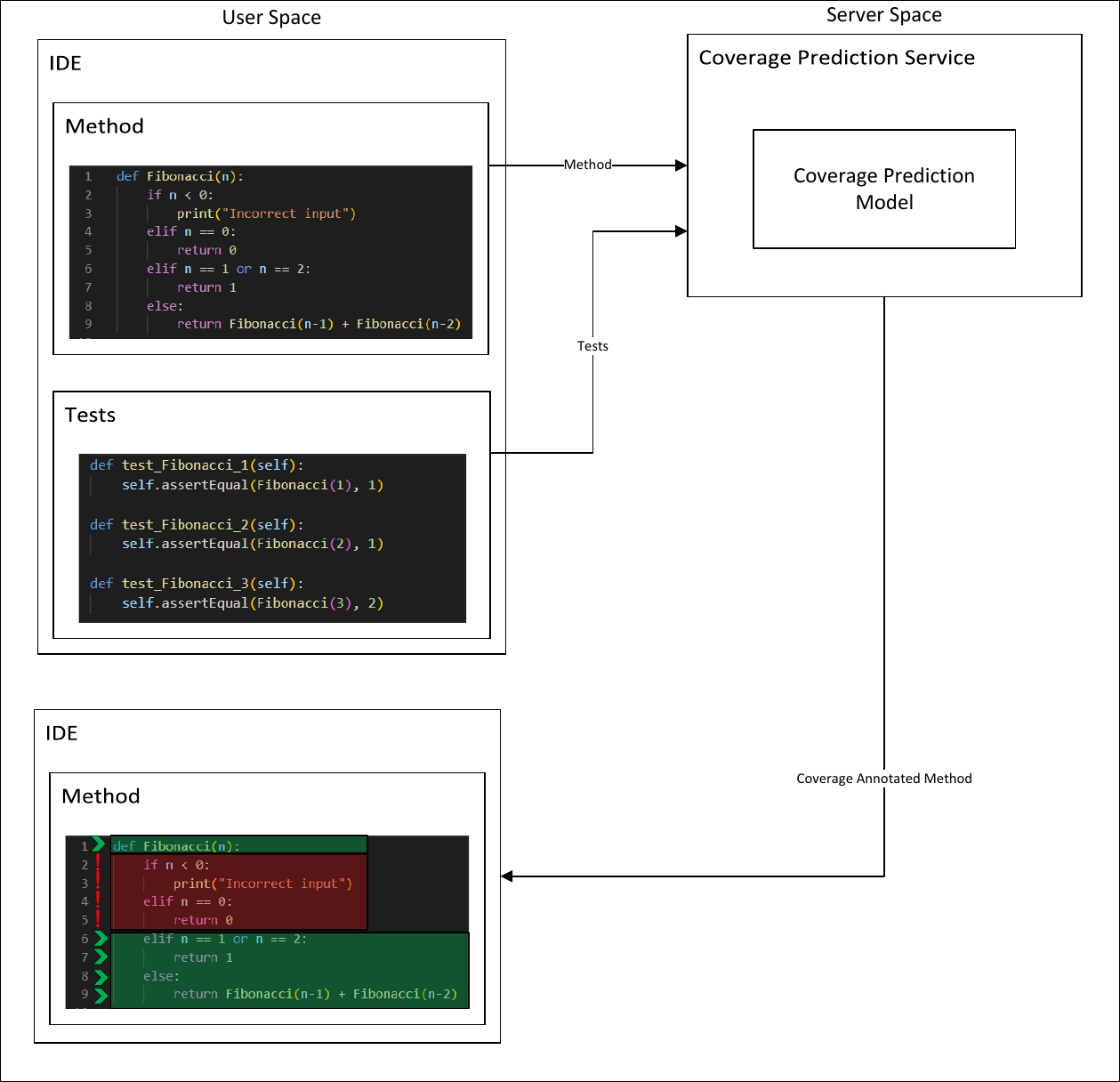}
    \caption{System A - Live Coverage}
    \label{fig:system_A}
\end{figure*}

Figure \ref{fig:system_A} shows the deployment of System A, which provides live coverage prediction for developers directly into their IDE. System A supports the scenario where a developer is writing tests for a given method (\eg \texttt{Fibonacci(n)}) in their codebase. System A provides live coverage information (bottom of Figure \ref{fig:system_A}) where lines covered by the tests are marked with \colorbox{green}{>} and highlighted in green and the line missed are marked with \colorbox{red}{!} and highlighted in red.

The benefits provided by System A are the following: (i) no need to build the entire codebase; (ii) no need to execute the tests; (iii) live and lightweight coverage prediction.

\subsection{System B - Test Generation with Coverage}
\begin{figure*}
\includegraphics[width=0.95\textwidth]{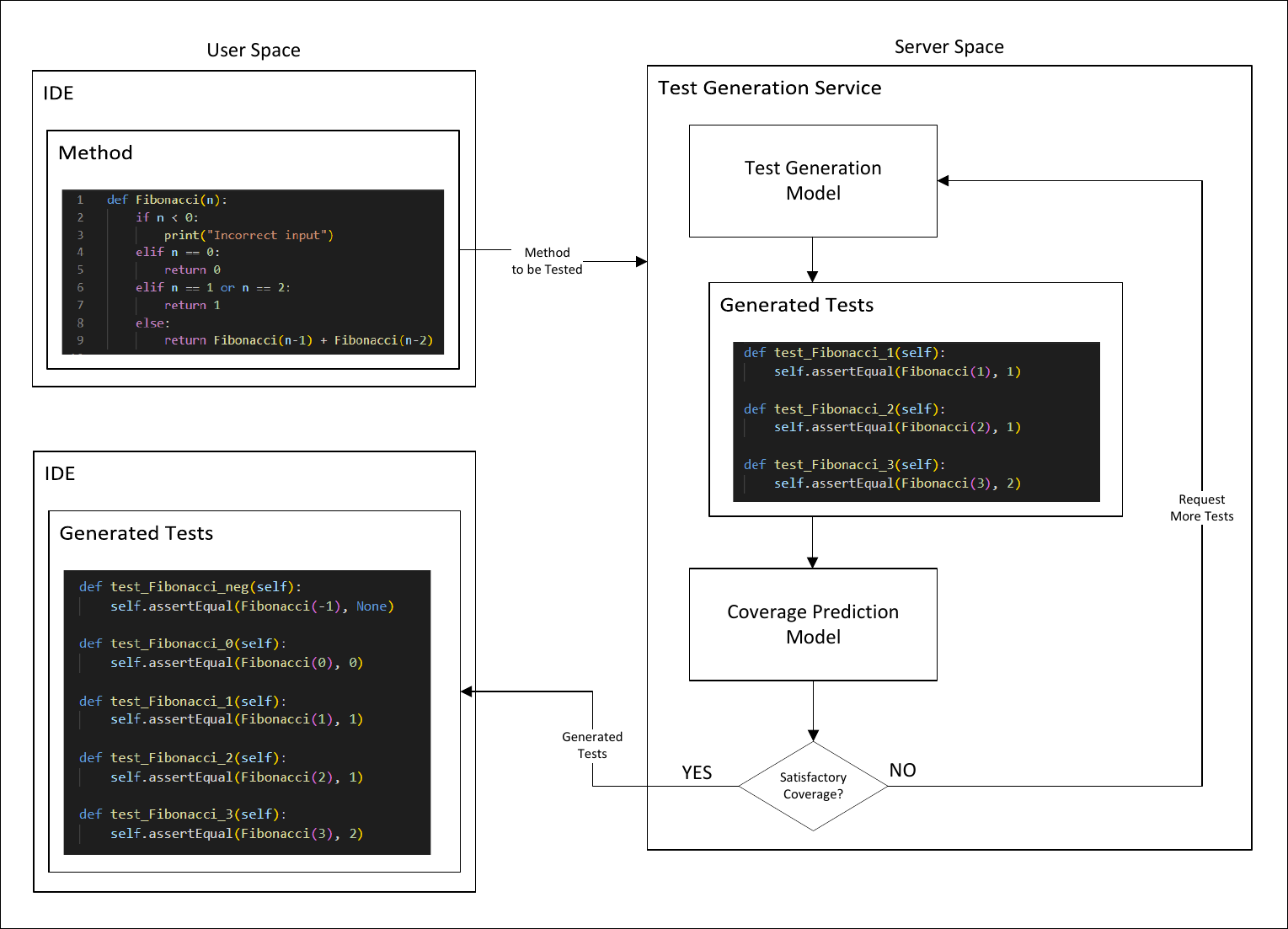}
    \caption{System B - Test Generation with Coverage}
    \label{fig:system_B}
\end{figure*}

Figure \ref{fig:system_B} shows the deployment of System B, which provides Test Suites with a coverage guarantee. System B supports the scenario where a developer is requesting test cases for a given method and would like to obtain a certain degree of coverage on the method under test. Once the method is transmitted to the Test Generation Service, the Test Generation Model (\ie an AI-based test generation tool or any other tool) outputs a first batch of test case candidates. The Coverage Prediction Model analyzes these tests and the method under test, and predicts the coverage that these tests achieve on the method. If the coverage is satisfactory (w.r.t. a given criteria and threshold) the tests are transmitted to the IDE and shown to the developer. If the tests do not meet the criteria in terms of coverage, the Test Generation Service requests additional tests from the Test Generation Model (optionally, providing the specific lines/branches which still need to be covered). 

The benefits provided by System B are the following: (i) automated test generation with coverage guarantees; (ii) lightweight generation without need of build and test execution on the user side.

\end{document}